\documentclass[prb,twocolumn,showpacs,superscriptaddress,amsmath]{revtex4}
\usepackage{graphicx}
\usepackage{amssymb}
\usepackage{citesort}

\begin{document}
\title{Cooperative orbital ordering and Peierls instability in the checkerboard lattice with doubly degenerate
orbitals}
\author{R.T. Clay}
\affiliation{Department of Physics and Astronomy and HPC$^2$ Center for Computational
 Sciences, Mississippi State University, Mississippi State MS 39762}
\author{H. Li}
\affiliation{ Department of Physics, University of Arizona, Tucson, AZ 85721}
\author{S. Sarkar}
\affiliation{S. N. Bose National Centre for Basic Sciences, Kolkata, India }
\author{S. Mazumdar}
\affiliation{ Department of Physics, University of Arizona, Tucson, AZ 85721}
\author{T. Saha-Dasgupta}
\affiliation{S. N. Bose National Centre for Basic Sciences, Kolkata, India }
\date{\today}
\begin{abstract}
 It has been suggested that the metal-insulator transitions in a
 number of spinel materials with partially-filled $t_{2g}$
 $d$-orbitals can be explained as orbitally-driven Peierls
 instabilities.  Motivated by these suggestions, we examine
 theoretically the possibility of formation of such orbitally-driven
 states within a simplified theoretical model, a two-dimensional
 checkerboard lattice with two directional metal orbitals per atomic
 site.  We include orbital ordering and inter-atom electron-phonon
 interactions self-consistently within a semi-classical approximation,
 and onsite intra- and inter-orbital electron-electron interactions at
 the Hartree-Fock level.  We find a stable, orbitally-induced Peierls
 bond-dimerized state for carrier concentration of one electron per
 atom.  The Peierls bond distortion pattern continues to be period 2
 bond-dimerization even when the charge density in the orbitals
 forming the one-dimensional band is significantly smaller than 1.  In
 contrast, for carrier density of half an electron per atom the
 Peierls instability is absent within one-electron theory as well as
 mean-field theory of electron-electron interactions, even for nearly
 complete orbital ordering.  We discuss the implications of our
 results in relation to complex charge, bond, and orbital-ordering
 found in spinels.
\end{abstract}

\maketitle

\section{Introduction}

Transition-metal spinel compounds AB$_2$X$_4$, where X is S or O, have
been for many years the subject of intense experimental and
theoretical activity.  The B sublattice of the spinel structure forms
corner-sharing tetrahedra, giving rise to a geometrically frustrated
pyrochlore lattice.  In general, oxides and chalcogenides of
transition metals exhibit complex charge- and spin-ordering that are
often coupled with orbital-ordering.  \cite{Tokura00a,Dagotto05b}
Coupled orbital-charge-spin orderings have been investigated widely
for compounds of the late transition metal ions with active $e_g$
orbitals within cubic geometry, such as the cuprates and manganites.
In contrast, such orderings in spinel compounds are only beginning to
be studied \cite{Radaelli05a}.  Spinel systems are considerably more
complicated than the cuprates or manganites due to two distinct
reasons.  First, the B-ions in the spinels, which are in an octahedral
environment of the X anions, often possess partially-filled
$t_{2g}$ $d$-orbitals. Examples include V$^{3+}$ ions in the vanadates
of Zn \cite{Tsunetsugu03a,Tsunetsugu05a}, Mn \cite{Sarkar09b}, and Cd
\cite{Nishiguchi02a}; V$^{2.5+}$ in AlV$_{2}$O$_{4}$
\cite{Matsuno01a}; Ti$^{3+}$ in MgTi$_{2}$O$_{4}$ \cite{Schmidt04a};
Ir$^{3.5+}$ in CuIr$_{2}$S$_{4}$ \cite{Radaelli02a}; and Rh$^{3.5+}$
in LiRh$_{2}$O$_{4}$ \cite{Okamoto08a}.  The threefold degeneracy of
the t$_{2g}$ orbitals, along with the weaker tendency to Jahn-Teller (JT)
distortions \cite{Dunitz57a} in $t_{2g}$-based systems lead to more
complicated physics compared to $e_g$-based systems.  Second, the
geometric frustration in the spinel structure, mentioned above,
precludes simple orderings \cite{Radaelli05a}.

Very recently, an exotic orbitally-induced Peierls instability has
been proposed as the mechanism behind the metal-insulator transitions
in the spinels like CuIr$_{2}$S$_{4}$ \cite{Khomskii05a}.
Charge-segregation of Ir into formally Ir$^{3+}$ and Ir$^{4+}$,
accompanied by the formation of an octamer of Ir$^{4+}$ ions with
alternate short and long bonds is observed in CuIr$_2$S$_4$
\cite{Radaelli02a}. This unusual charge-ordering pattern has been
explained within the context of an effective one-dimensionalization
and Peierls instability driven by orbital ordering (OO). The spinel
structure of CuIr$_{2}$S$_{4}$ consists of criss-cross chains of
Ir-ions with strongest overlaps between orbitals of the same kind
(i.e., $d_{xy}$ with $d_{xy}$, $d_{yz}$ with $d_{yz}$).  Within the
proposed mechanism, OO in Ir$^{3.5+}$ ions leads to completely filled
degenerate $d_{xz}$ and $d_{yz}$ orbitals and effective
one-dimensional (1D) quarter-filled $d_{xy}$ bands. The latter now
undergoes a Peierls instability that is accompanied by period 4
charge-ordering $\cdots$ Ir$^{3+}$/Ir$^{3+}$/Ir$^{4+}$/Ir$^{4+}$
$\cdots$ and long-intermediate-short-intermediate bonds, as seen
experimentally\cite{Radaelli02a}.  The OO and Peierls instability are
thought to be coupled, as opposed to independent. Closely related
phenomenologies have been proposed to explain the metal-insulators
transitions in MgTi$_{2}$O$_{4}$ \cite{Khomskii05a} and
LiRh$_{2}$O$_{4}$ \cite{Okamoto08a}.

While the proposed scenario \cite{Khomskii05a} does provide
qualitative explanations, it raises many interesting questions.  For
example, the bond or charge periodicities that result from Peierls
distortion in real quasi-1D systems depend strongly on the
bandfilling.  Naively then, one might be led to believe that unless
the OO is complete and leads to integer occupations of the individual
$t_{2g}$ orbitals, exactly commensurate distortion periodicities, as
supposedly observed in CuIr$_{2}$S$_{4}$ \cite{Radaelli02a} and
MgTi$_{2}$O$_{4}$ \cite{Schmidt04a}, are not expected.  The exactly
period 4 distortion in CuIr$_{2}$S$_{4}$ and MgTi$_{2}$O$_{4}$ is
therefore a puzzle. Secondly, the role of electron-electron (e-e)
interactions, which can have important consequences on these
transitions, is not clear. For example, it has been suggested that the
bond distortion in MgTi$_{2}$O$_{4}$ causes nearest-neighbor Ti$^{3+}$
pairs to form a spin-singlet state, giving a drop in the magnetic
susceptibility at low temperature\cite{Isobe02a}.  Furthermore, e-e
interactions can strongly affect the distortion periodicities in 1D
chains with fillings different from 1 electron per atom
\cite{Mazumdar84a}.  Addressing these issues requires explicit
calculations based on Hamiltonians containing all the necessary
components, that to the best of our knowledge do not exist currently.

We report here the results of explicit calculations based on a
simplified model system that displays co-operative OO and Peierls
instability.  Our model system is a two-dimensional projection of the
pyrochlore lattice, a checkerboard lattice with two degenerate
directional orbitals per atom.  As we discuss in the next Section, the
model captures the effective one-dimensionalization that occurs in the
spinel lattice. There exist also subtle differences between the our
model and the theoretical picture that has been employed for the
spinels \cite{Khomskii05a}, that we discuss in Section
\ref{sect_discussion}.  We have considered two different electron
densities, 1 electron per atom and $\frac{1}{2}$ electron per atom.
We find that a stable Peierls state with 1D order occurs for an
electron density of 1 electron per atom. Importantly, even for
significant deviations of orbital occupancies from integer values, the
Peierls distortion is purely bond dimerization, as would be true in
the precisely $\frac{1}{2}$-filled 1D band.  We speculate that this is
a consequence of interactions between the effective 1D chains of our
model, which are not entirely independent.  For the case of
$\frac{1}{2}$ electron per atom we did not observe an orbitally-driven
Peierls state even for integer orbital occupancies in the absence of
e-e interactions. For the same model in the limit of infinite e-e
interactions, however, a stable bond dimerization occurs. This
suggests the second main result of our paper, namely that e-e
interactions are important in stabilizing orbitally-induced Peierls
states for the case of $\frac{1}{2}$-electron per atom within the
orbitally degenerate checkerboard lattice.  We will discuss possible
implications of this result for the spinel lattice.

The paper is divided in following subsections: in Section
\ref{sect_model} we introduce our model Hamiltonian; in Sections
\ref{sect_1e} and \ref{sect_12e} we present numerical results for 1
and $\frac{1}{2}$ electron per atom respectively; and in Section
\ref{sect_discussion} we conclude and further discuss the relationship
of our theory to spinels and related materials.
\begin{figure}
\centerline{\resizebox{3.0in}{!}{\includegraphics{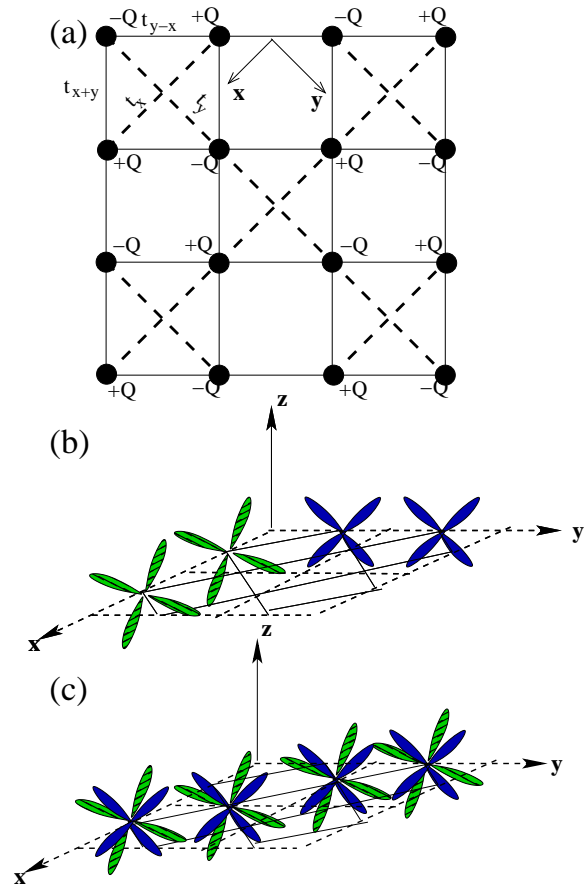}}}
\caption{(color online) (a) 2D checkerboard lattice with
  orbital-ordering pattern given by the $H_{\rm{OO}}$ term in
  Eq.~\ref{ooham}.  Filled dots denote metal atom positions, with two
  orbitals per atom, $xz$ and $yz$.  Hopping terms are included along
  $x$, $y$, $x+y$, and $y-x$ directions as defined in
  Eq.~\ref{ty-x}--Eq.~\ref{ty}.  Hopping along the $t_x$ and $t_y$
  directions, indicated by dashed lines, is modulated by the
  inter-atomic e-p coupling. (b) The $(dd\pi)$ overlap of $xz$, shown
  in green (light grey), and $yz$, shown in blue (dark grey), along
  the x- and y-directions. Note that along these directions, non-zero
  hopping matrix elements exist only along $x$- ($xz$-$xz$) or $y$-
  ($yz$-$yz$) direction. (c) The $(dd\pi)$ overlap of $xz$ and $yz$
  orbitals along the $y-x$ direction. The overlap along $x+y$ is
  similar apart from a change of sign for the inter-orbital terms.}
\label{lattice}
\end{figure}     

\section{Theoretical model}
\label{sect_model}

As discussed above, no quantitative calculations exist of coexisting
orbital and Peierls order in spinel lattices. This is due to the huge
complexity of these systems, which include three dimensionality,
triply degenerate $t_{2g}$ metal atom orbitals, possible JT
distortions and frustration, in addition to e-e interactions.  We
therefore construct a simpler model based on a checkerboard lattice
that is easier to handle, but that is expected to bear similarities
with the spinel problem.  Note that although there exists a large body
of literature on the consequences of frustration within the
checkerboard lattice with a {\it single} atomic orbital per site
\cite{Fouet03a,Tchernyshyov03a,Runge04a,Starykh05a,Poilblanc07a,Trousselet08a,Yoshioka08a,Yoshioka08b},
we are unaware of similar calculations on the present model which
deals with doubly degenerate orbitals at each site.  An important
property of orbitally-driven Peierls order in these materials is that
the OO leads to Peierls order in {\it several different crystal
  lattice directions} \cite{Khomskii05a}.  In order to incorporate
orbital degeneracy, frustration, and the possibility of Peierls order
in multiple directions we consider the following Hamiltonian for a
checkerboard lattice with doubly degenerate metal orbitals at each
lattice site.
\begin{eqnarray}
H&=&H_{SSH}+H_{OO}+H_{ee} \label{spinelham} \\
H_{SSH}& = & \sum_{i,\bf{a},\gamma,\gamma^\prime} 
t^{\bf{a}}_{\gamma\gamma^\prime}(1 + \alpha_{\bf a}
\Delta_{i,i+{\bf a}}) (d^\dagger_{i\gamma\sigma}d_{i+\bf{a}\gamma^\prime\sigma}\nonumber \\
&+&h.c.)  +\frac{1}{2}\sum_{i\bf{a}} K_{\rm{SSH}} \Delta_{i,i+{\bf a}}^2 \label{sshham} \\  
H_{OO}& =& \frac{g}{2} \sum_{i,\gamma \neq \gamma^\prime} Q_i (n_{i\gamma^\prime} - n_{i\gamma}) + 
\frac{1}{2}K_{OO}\sum_i Q_i^2 \label{ooham} \\
H_{ee}& =& U\sum_{i,\gamma} n_{i\gamma\uparrow}n_{i\gamma\downarrow} +
\frac{U^\prime}{2}\sum_{i,\gamma\neq\gamma^\prime}n_{i\gamma}n_{i\gamma^\prime} \label{hubbham}
\end{eqnarray}

The Hamiltonian in Eq.~\ref{spinelham} consists of, (i) $H_{SSH}$ that
contains the kinetic energy and the inter-ion electron-phonon (e-p)
coupling (Eq.~\ref{sshham}), (ii) an OO term $H_{OO}$
(Eq.~\ref{ooham}), and (iii) e-e interaction $H_{ee}$
(Eq.~\ref{hubbham}) that includes short-range e-e interactions within
each site.  We describe each of these terms separately below.

$H_{SSH}$ includes electron hopping between same as well as different
orbitals.  In Eq.~\ref{sshham}, $d^\dagger_{i\gamma\sigma}$ creates an
electron of spin $\sigma$ in the orbital $\gamma$ of atom $i$.
$\gamma$ and $\gamma^{\prime}$ correspond to $d_{xz}$ and $d_{yz}$
orbitals, which have lobes that are oriented perpendicular to each
other at each site as shown in Fig.~\ref{lattice}.  The inter- and
intra-orbital hopping matrix elements $t^{\bf
  a}_{i\gamma\gamma^\prime}$ are based on Slater-Koster
parametrization of hopping integrals connecting $t_{2g}$ orbitals
\cite{HarrisonBook}.  {\bf a} denotes a unit vector along the $x$,
$y$, $x+y$, or $y-x$ directions.  Each bond indicated in
Fig.~\ref{lattice} connects two orbitals at each metal ion site with
two orbitals at another site, and is hence written as a 2$\times$2
matrix:
\begin{eqnarray}
 \left( \begin{array}{cc} t_{xz,xz} & t_{xz,yz} \\ t_{yz,xz} & t_{yz,yz} \\ \end{array} \right)_{y-x}& =&
 \left(  \begin{array}{cc} -\frac{1}{2} &-\frac{1}{2} \\ -\frac{1}{2}&  -\frac{1}{2} \\  \end{array} \right) \label{ty-x} \\
 \left( \begin{array}{cc} t_{xz,xz} & t_{xz,yz} \\ t_{yz,xz} & t_{yz,yz} \\ \end{array} \right)_{x+y}& = &
 \left(  \begin{array}{cc} -\frac{1}{2} &\frac{1}{2} \\ \frac{1}{2}&  -\frac{1}{2} \\  \end{array} \right) \label{txy}  \\
\left( \begin{array}{cc} t_{xz,xz} & t_{xz,yz} \\ t_{yz,xz} & t_{yz,yz} \\ \end{array} \right)_{x}& =&
 \left(  \begin{array}{cc} -1 &0 \\ 0&  0 \\  \end{array} \right)  \label{tx} \\
\left( \begin{array}{cc} t_{xz,xz} & t_{xz,yz} \\ t_{yz,xz} & t_{yz,yz} \\ \end{array} \right)_{y}& =&
 \left(  \begin{array}{cc} 0 &0 \\ 0& -1 \\  \end{array} \right) \label{ty}
\end{eqnarray}    
All hopping integrals are in units of the $(dd\pi)$ matrix element
(set to -1) involving the two $d$-orbitals of the metal atoms and
mediated by the $p$-orbital of the anion (not shown explicitly in the
figure) in between. In the above, the small $dd\delta$ hoppings have
been neglected.  Note that electron hoppings along the $x$ ($y$)
direction involve only the $d_{xz}$ ($d_{yz}$) orbitals. Thus the
$\pi$-bonding among the orbitals, as opposed to the $\sigma$-bonding
in spinels, does not preclude the orbitally-induced effective
one-dimensionalization \cite{footnote}.

The inter-ion e-p coupling in $H_{SSH}$ is written in the usual
Su-Schrieffer-Heeger (SSH) form \cite{Su80a}. Here $\alpha_{\bf a}$ is
the e-p coupling constant corresponding to the bond between atoms at
$i$ and $i+{\bf a}$, $\Delta_{i,i+{\bf a}}$ the deviation of this bond
from its equilibrium length, and $K_{\rm{SSH}}$ the corresponding
spring constant. We include nonzero e-p couplings only along the $x$
and $y$-directions, in keeping with the Peierls distortions in
CuIr$_2$S$_4$ and MgTi$_2$O$_4$ involving only orbitals of the same
kind, and assume e-p couplings of equal strength ($\alpha_x = \alpha_y
= \alpha$).  We take all spring constants $K_{\rm{SSH}}=1$.
\begin{figure}
\centerline{\resizebox{2.5in}{!}{\includegraphics{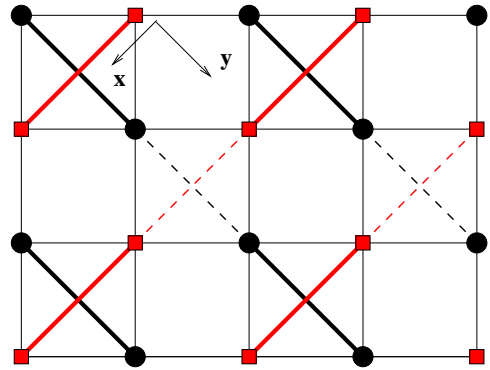}}}
\caption{(color online) Pattern of the orbital-ordering and bond-order
  modulation for one electron per atom. Squares (circles) represent
  atoms with predominant xz (yz) orbital occupation. Bonds alternate
  in strength along diagonal directions, with solid (dashed) lines
  indicating strong (weak) bonds.}
\label{dist-1e}
\end{figure}

The OO term $H_{OO}$ breaks the orbital degeneracy between the two
orbitals at each individual site.
$n_{i\gamma\sigma}=d^\dagger_{i\gamma\sigma}d_{i\gamma\sigma}$ is the
number of electrons of spin $\sigma$ on orbital $\gamma$ of site $i$,
and $n_{i\gamma}=n_{i\gamma\uparrow}+n_{i\gamma\downarrow}$.  The
coordinate $Q_i$ couples to the charge density difference between the
two orbitals on the site, with a corresponding coupling constant $g$
and spring constant $K_{OO}$. We fix $K_{OO}$ to the value 1.  As
written in Eq.~\ref{ooham}, the relative phase of the $Q_i$ at each
site is unrestricted.  Within the model of reference
\onlinecite{Khomskii05a} the effective one-dimensionalization is a
consequence of OO alone and in principle will occur even in the
absence of Peierls distortion. We have used this to determine the
preferred $Q_i$ mode for both 1 and $\frac{1}{2}$ electron per site by
calculating the orbital occupations in several large lattices with
open boundary condition (OBC).  In a large OBC lattice, the charge
densities and bond orders far from the lattice edges spontaneously
assume the pattern that would occur in the infinite lattice for 0$^+$
coupling limit \cite{Mazumdar00a}. An alternate approach to determine
the dominant OO mode is to calculate in a periodic lattice the
energies corresponding to each mode for fixed distortion amplitude;
the dominant mode is simply the one with the lowest total energy.  In
the present case we have performed both sets of calculations for both
1 and $\frac{1}{2}$ electron per site and have determined that the
preferred OO mode in both cases is the ``checkerboard'' pattern of
Fig.~\ref{lattice}(a), which can be parametrized in terms of a single
amplitude $|Q|$ with $Q_i=(-1)^{i_x+i_y}|Q|$, where $i_x$ and $i_y$
are the $x$ and $y$ coordinates of the $i$th atom.
   
The third term in the Hamiltonian, $H_{ee}$ (Eq.~\ref{hubbham}),
includes short-ranged Coulomb repulsions.  $U$ $(U^\prime)$ is the
on-atom Coulomb repulsion for electrons in same (different) orbitals.
While exact diagonalization has been used successfully for many 1D,
quasi-1D, and two dimensional (2D) lattices involving both e-e and e-p
interactions, in the present model with two orbitals per metal site
the Hilbert space is too large to treat any meaningful size cluster
within exact diagonalization.  We will therefore consider first the
non-interacting ($U=U^\prime=0$) system, and then consider the effect
of $U$ and $U^\prime$ within the unrestricted Hartree-Fock (UHF)
approximation, with no assumption of the periodicity of the UHF
wavefunction.  Specifically, we replace the interaction terms in
$H_{ee}$ (Eq.~\ref{hubbham}) by
\begin{eqnarray*}
&&U\sum_{i,\gamma, \sigma, \sigma^\prime}n_{i\gamma\sigma}\langle n_{i\gamma\sigma^\prime} \rangle
 - U\sum_{i,\gamma} \langle n_{i\gamma\downarrow} \rangle
\langle n_{i\gamma\uparrow} \rangle \nonumber \\
&+& \frac{U^\prime}{2}[\sum_{i,\gamma\neq\gamma^\prime,\sigma, \sigma^\prime} (n_{i\gamma\sigma}(\langle 
n_{i\gamma^\prime\sigma}\rangle + \langle n_{i\gamma^\prime\sigma^\prime} \rangle) \nonumber \\
&-& (d^\dagger_{i\gamma^\prime\sigma}d_{i\gamma\sigma}+H.c.)\langle
 d^\dagger_{i\gamma^\prime\sigma}d_{i\gamma\sigma}+H.c.\rangle \nonumber \\
&-& \langle n_{i\gamma\sigma} \rangle \langle n_{i\gamma^\prime\sigma} \rangle -
\langle n_{i\gamma\sigma} \rangle \langle n_{i\gamma^\prime\sigma^\prime} \rangle 
+ \langle  d^\dagger_{i\gamma^\prime\sigma}d_{i\gamma\sigma}+H.c.\rangle^2] 
\end{eqnarray*}
$\langle n_{i\gamma\sigma} \rangle$ and $\langle
d^\dagger_{i\gamma^\prime\sigma}d_{i\gamma\sigma}+H.c.\rangle$ are
obtained using a combination of self-consistency and simulated
annealing for finding their ground state values.  The UHF
approximation often gives unphysical results for large
interaction strengths and we will primarily focus on small
$U$ and $U^\prime$.

We treat the OO and e-p interactions using a standard self-consistent
approach derived from the equations
\begin{equation}
\frac{\partial \langle H \rangle}{\partial Q} =0 \qquad
\frac{\partial \langle H \rangle}{\partial \Delta_{i,i+{\bf a}}} = 0. \label{selfconsist}
\end{equation}
The self-consistency equations derived from Eq.~\ref{selfconsist} are
used iteratively given an initial starting distortion. In the infinite
system the OO or the bond distortion would occur for
infinitesimally small coupling constants $g$ and $\alpha$.  In
finite-size clusters, however, due to the finite-size gaps between
successive energy levels, nonzero coupling constants are required
before the symmetry-broken state appears.  In the following we
consider $g$ and $\alpha$ close to the minimum values needed for the
broken-symmetry state to occur.

We performed calculations for lattices up to 16$\times$16 (256 atoms
with 512 orbitals). The primary difficulty in solving
Eq.~\ref{selfconsist} is that because of the large number of quantities
($Q$, $\Delta_{i,i+{\bf a}}$, and UHF average charge densities) to be
determined self-consistently, the calculations often became trapped in
local minima before reaching the true ground state. In all cases we
have taken care that the true ground state was reached.  Below we
summarize our numerical results. These are divided into two
subsections that discuss average charge densities of 1 electron per
atom and $\frac{1}{2}$ electrons per atom, respectively.

\section{Results}
\label{sect_results}

\subsection{1 electron per atom}
\label{results_1e}
\label{sect_1e}

\begin{figure}
\centerline{\resizebox{3.0in}{!}{\includegraphics{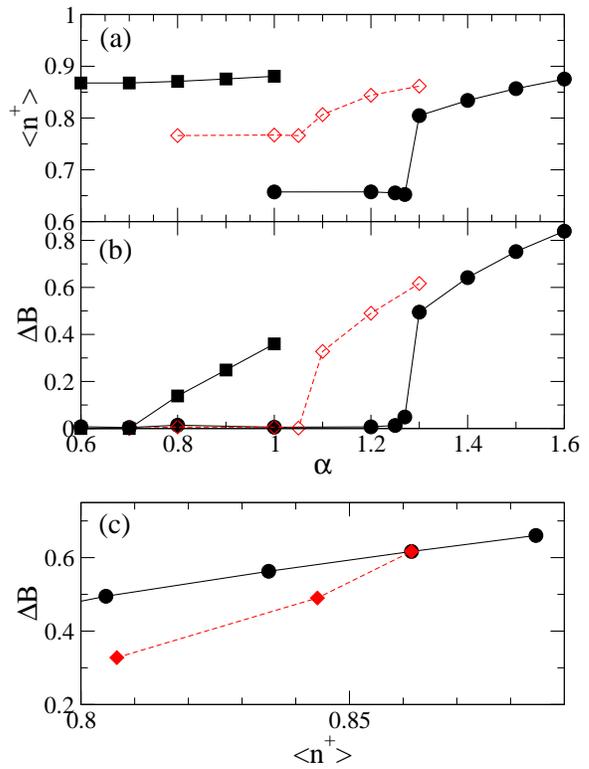}}}
\caption{(color online) Co-operative orbital-ordering and Peierls bond
  alternation for one electron per atom, with
  $U=U^\prime=0$. Calculations are for a 16$\times$16 periodic
  lattice.  (a) Majority charge density (see text) in the orbitals forming the
  quasi-1D chains following orbital ordering. Here and in (b) circles,
  diamonds and squares correspond to $g$ = 0.6, 0.8 and 1.0,
  respectively.  (b) $\Delta B$ along the diagonal chain directions
  (see text and Fig.~\ref{dist-1e}) as a function of $\alpha$.  (c) Bond
  alternation plotted as a function of charge density.  Circles show
  the effect of increasing $g$ with constant $\alpha$=1.3.  Diamonds
  show effect of increasing $\alpha$ with constant $g$=0.8. Lines are
  guides to the eye.}
\label{half-u0}
\end{figure}
The OO term in our model makes the orbital occupancy of the $xz$ and
$yz$ orbitals unequal at each site. To measure the degree of OO
quantitatively, we calculate the majority charge density $\langle
n^+\rangle$, defined as the charge density in the $xz$ orbitals at
$+|Q|$ sites (see Fig.~\ref{lattice}(a)).  These orbitals form
quasi-1D chains in the $x$ direction (sites denoted by squares in
Fig.~\ref{dist-1e}).  With one electron per site, $\langle n^+\rangle$
ranges from 0.5 to 1, with $\langle n^+\rangle=1$ indicating complete
OO, and $\langle n^+\rangle=0.5$ implying the absence of OO.  Due to
the symmetries present in our model, the quasi-1D chains along $x$ and
$y$ directions are identical--the charge density in the $yz$ orbitals
at $-|Q|$ sites (denoted by circles in Fig.~\ref{dist-1e}) is also
$\langle n^+\rangle$.

As an order parameter for the Peierls distortion, we measure the
modulation of the bond order, 
\begin{equation}
B_{i,i+{\bf a},\gamma}=\sum_\sigma \langle d^\dagger_{i+{\bf a},\sigma,\gamma}d_{i,\sigma,\gamma}
+H.c. \rangle.\label{bondorder}
\end{equation}
The bond order we are interested in (Eq.~\ref{bondorder}) is the
expectation value of charge-transfer between orbitals of same symmetry
belonging to neighboring atoms.  The charge-transfer is directly
coupled to the bond distortion $\Delta_{i,i+{\bf a}}$ in
Eq.~\ref{sshham}, and hence for nonzero $\Delta_{i,i+{\bf a}}$ the
charge transfer across consecutive bonds shows periodic modulation. The
extent of modulation of $B_{i,i+{\bf a},\gamma}$ is therefore a direct
measure of the SSH distortion strength.  As shown in
Fig.~\ref{dist-1e}, we find bond order modulation along $x$ ($y$)
direction to involve $xz$ ($yz$) orbitals only.  The modulation is
purely period-2 (dimerization) with alternating strong and weak bonds,
and hence we use $\Delta B$, the {\it difference} between the
calculated strong and weak bond orders involving orbitals of a
particular symmetry, as the order parameter for the SSH
distortion. Because of symmetry, the amplitudes of the bond order
modulations involving the $xz$ orbitals along the $x$-direction, and
the $yz$ orbitals along the $y$-directions are identical.

As discussed above, whether or not a co-operative orbitally-induced
Peierls instability occurs for $\langle n^+ \rangle < 1$, as well as
the periodicity of the resultant bond order wave are both important
issues.  We first consider the non-interacting limit ($U=U^\prime=0$).
The co-operative nature of the OO and bond dimerization is shown in
Fig.~\ref{half-u0}, where we show the results of our self-consistent
calculations for 16$\times$16 lattices. As seen in
Figs.~\ref{half-u0}(a) and (b), the orbitally-induced Peierls state
appears only for $\langle n^+ \rangle \geq 0.8$, with the bond
distortion a pure bond dimerization regardless of $\langle n^+
\rangle$.  As expected for a co-operative transition, $\langle n^+
\rangle$ increases with $g$, as seen in Fig.~\ref{half-u0}(a).  For
each $g$ there exists an $\alpha_c$ beyond which there occur
simultaneous jumps in $\langle n^+ \rangle$ and $\Delta B$ (the jump
in $\langle n^+ \rangle$ becomes progressively smaller as $g$
increases.)  The magnitude of $\alpha_c$ decreases with increasing $g$
(see Fig.~\ref{half-u0}(b)) To further show the cooperative effect, in
Fig.~\ref{half-u0}(c) we show the effects of (i) increasing $\alpha$
at constant $g$, and (ii) increasing $g$ at constant $\alpha$.  While
it is to be expected that the orbital order parameter $\langle n^+
\rangle$ increases with $g$, or that the bond alternation parameter
increases with $\alpha$, we find that either of the coupling constants
enhances both $\langle n^+ \rangle$ and $\Delta B$.
\begin{figure}
\centerline{\resizebox{3.0in}{!}{\includegraphics{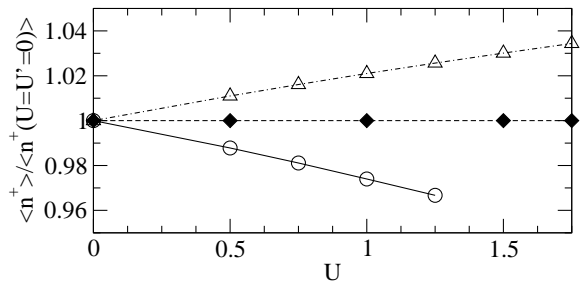}}}
\caption{Majority charge density as a function of $U$ and $U^\prime$
  for one electron per atom, normalized with respect to the same
  quantity for the uncorrelated system.  Results shown are for for
  16$\times$16 periodic lattices with $g=0.8$, $\alpha=1.2$, and
  $K_{OO}=K_{SSH}=1$.  Circles are for $U^\prime=0.3U$, diamonds are
  for $U^\prime=0.5U$, and triangles are for $U^\prime=0.7U$.  Lines
  are guides to the eye.}
\label{half-u}
\end{figure}

Next we consider the correlated case with nonzero $U$ and $U^\prime$.
As for $U=U^\prime=0$, the orbitally-driven Peierls state is again
bond-dimerized.  In Fig.~\ref{half-u} we plot $\langle n^+ \rangle$
normalized by its value in the uncorrelated system as a function of
$U$ for several values of $U^\prime$. Within UHF, the
combined effect of $U$ and $U^\prime$ can be to either weaken or
strengthen the distortion: for fixed $U^\prime$, $U$ tends to weaken
the OO and the bond distortion, while for fixed $U$,
$U^\prime$ strengthens both order parameters. Within the UHF
approximation for one electron per atom, the effects of $U$ and
$U^\prime$ cancel exactly when $U^\prime=\frac{1}{2}U$.

From the Hamiltonian, the consequence of $U^\prime$ is to minimize the
intrasite inter-orbital Coulomb repulsion, which is achieved by orbital
ordering.  It is thus not surprising that $U^\prime$ has the same
effect as $g$ in Fig.~\ref{half-u}. The effect of $U$, as seen in
Fig.~\ref{half-u}, is however an artifact of the UHF
approximation. In the case of the strictly 1D $\frac{1}{2}$-filled band chain
with one orbital per site, exact diagonalization and quantum Monte
Carlo calculations have shown that the Peierls bond-alternation is
{\it enhanced} by the Hubbard $U$ \cite{Baeriswyl92a}. In contrast, the
UHF approximation predicts incorrectly that $U$ destroys the bond alternation
in the above case \cite{Baeriswyl92a}.  Had we been able to perform
exact diagonalization in the present case, we would have found similar
enhancement of the bond dimerization by $U$. This would have had a
profound effect on our overall result, reducing significantly the
$\alpha_c$ or the threshold $\langle n^+ \rangle$ at which the bond
dimerization appears.

The most important conclusion that follows from the above is that an
orbitally-induced Peierls instability can occur even for incomplete OO
($\langle n^+ \rangle \sim 0.8$), and as long as the instability
occurs at all, the bond order wave is period 2 for $\langle n^+
\rangle$ significantly less than 1. Indeed, it is conceivable that the
threshold value of $\langle n^+ \rangle$ at which the bond
dimerization appears can be even smaller than 0.8 for nonzero e-e
interactions. We have found no other periodicity or evidence for
soliton formation in our calculations.  An interesting aspect of the
OO driven bond distortion here is the phase relationship between the
bond order wave states involving the $d_{xz}$ and $d_{yz}$ orbitals in
Fig.~\ref{dist-1e}. The short bonds along the $x$ and $y$-directions
occupy the same plaquettes, yielding a structure that is reminiscent
of (but different from) the valence bond crystal obtained within the
Heisenberg spin-Hamiltonian for the checkerboard lattice
\cite{Fouet03a}.  Furthermore, the bond dimerizations along any one
direction but on different diagonals of the checkerboard lattice are
strictly ``in-phase''. Both of these indicate that while the bond
dimerizations are consequences of effective one-dimensionalization,
there exist strong 2D interactions in between both the criss-cross and
parallel chains. We ascribe the persistence of the bond dimerization
for $\langle n^+ \rangle < 1$ to the commensurability effect arising
from the 2D interactions.

\subsection{$\frac{1}{2}$ electron per atom}
\label{sect_12e}

\begin{figure}
\centerline{\resizebox{3.0in}{!}{\includegraphics{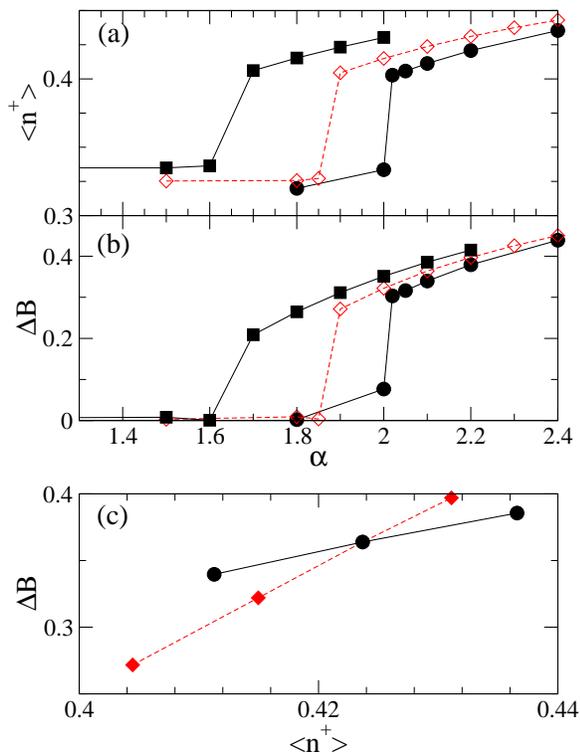}}}
\caption{(color online) Co-operative orbital-ordering and Peierls
  bond-alternation for 0.5 spinless fermions per atom. Calculations
  are for a 16$\times$16 periodic lattice.  (a) Majority charge
  density in the orbitals forming the 1D chains.  Here and in (b)
  circles, diamonds and squares correspond to $g$ = 0.6, 0.8 and 1.0,
  respectively.  (b) $\Delta B$ along the diagonal chain directions
  (see Fig.~\ref{dist-1e}) as a function of $\alpha$. (c) Bond
  alternation versus majority charge density.  Circles show the effect
  of increasing $g$ with constant $\alpha$=1.3.  Diamonds show effect
  of increasing $\alpha$ with constant $g$=0.8.  Lines are guides to
  the eye.}
\label{spinless}
\end{figure}
In CuIr$_2$S$_4$ and MgTi$_{2}$O$_{4}$, the distortion along the chain
directions is not bond dimerization but a period 4 distortion
\cite{Radaelli05a}.  This is as expected for a Peierls transition in a
1D chain with carrier density 0.5. We have therefore performed
self-consistent calculations within our model Hamiltonian also for
density 0.5.

Not surprisingly, we do obtain self-consistently an orbitally ordered
state here for nonzero $g$ with $\alpha=0$. Even within an essentially
100\% orbitally-ordered state (minority orbital charge density $\alt$
0.01), however, and with very strong $\alpha$, we were unable to
obtain a stable Peierls-distorted state. In all cases our
self-consistent simulations converged to states with disordered bond
distortions and charge densities, indicating vanishing bond-charge
distortion in the thermodynamic limit.  We obtained similar results
after including $U$ and/or $U^\prime$ within the UHF approximation. We
conclude that for electron density away from 1 electron per atom,
orbitally-driven Peierls ordering does not occur within our model in
the non-interacting limit or within the mean-field approach to e-e
interactions.

We ascribe the absence of bond-charge distortion here to the 
important role played by the interaction among the crisscrossing
chains within the checkerboard lattice. We have already pointed out in
Section \ref{sect_results}A that the stabilization of the perfect
period 2 distortion for the case of 1 electron per atom, even in the
absence of complete OO, is a signature of such a 2D
interaction. Similar 2D interactions should be relevant also for
carrier density 0.5. Since within mean field theory the Peierls
instability in 2D is limited to carrier density of 1, the absence of
the bond-charge distortion in the present case is to be
anticipated. On the other hand, we have recently shown in a series of
papers that specifically for this carrier density and one orbital per
site, nonzero e-e interactions can strongly stabilize bond-charge
ordered states in 2D lattices \cite{Mazumdar00a,Clay02a}. In the case
of the checkerboard lattice with a single orbital per site, plaquette
spin-singlet formation (though without charge-ordering) has similarly
been found for the same carrier concentration \cite{Indergand07a}.  It
is conceivable that similar effects of e-e interactions persist in the
present case with two degenerate orbitals per site.

Unfortunately, performing a realistic calculation with finite $U$ that
goes beyond the UHF approximation in the present case is beyond our
computational capability. We have therefore investigated our model
Hamiltonian Eq.~\ref{spinelham} in the limit of $U\rightarrow\infty$,
where we assume band orbital occupancy corresponding to that for
spinless fermions.  Fig.~\ref{spinless} shows the same order
parameters as in Fig.~\ref{half-u0} for the spinless fermion
case. While the transition occurs here for a slightly larger value of
the coupling constant $\alpha$, it is otherwise identical to the
transition with 1 electron per atom, viz., bond dimerization occurs
along the diagonal directions, and OO and bond
distortion reinforce each other cooperatively.

\section{Discussion}
\label{sect_discussion}

In summary, we have carried out numerical studies on the 2D
checkerboard lattice with two degenerate directional orbitals per site
- a model system that like the spinel compounds can in principle
exhibit OO-driven Peierls bond distortions and charge ordering in
multiple directions.  In addition to OO and bond modulation terms, our
model Hamiltonian includes both intra- and inter-orbital e-e
correlations that were treated within the UHF approximation.  Although
some of our results have strong implications for the spinels, it is
useful to precisely understand the differences between the spinel and
checkerboard lattices such that the applicability as well as
limitations of our model can both be understood.  One difference
between the two lattices is that the plaquettes in the 2D checkerboard
lattice do not correspond to the tetrahedra in the spinel lattice
because of the difference between horizontal and vertical bonds in
Fig.~\ref{lattice}(a) on the one hand and the diagonal bonds on the
other \cite{Fouet03a}. What is more important in the present context
is that the OO in our model is not driven by a band JT transition that
destroys the degeneracies of the atomic orbitals in the model of
\onlinecite{Khomskii05a}. Within our model the two orbitals of
different symmetries on a given atom are both potentially active
orbitals.

We believe that our demonstration of the co-operative interaction
between OO and the Peierls instability in Section \ref{sect_1e}, where
each broken symmetry enhances the other, is of direct relevance to the
$t_{2g}$-based spinel systems, where qualitative discussions have
suggested similar results \cite{Khomskii05a}.  Similarly, our
observation that the period 2 bond distortion persists for 1 electron
per atomic site even for incomplete OO, with majority charge density
as low as 0.8 per orbital, may also be of significance for the
spinels. This should be particularly true for nonzero onsite Hubbard
interaction, which will tend to decrease the amplitude of the OO.
Complete OO in the real systems CuIr$_2$S$_4$ and MgTi$_2$O$_4$
requires that the energy gap due to the JT distortion is significantly
larger than the Hubbard interaction.  It is at least equally likely
that the commensurate charge and bond distortions found in the
experimental systems are not due to complete OO but are consequences
of the complex interactions between the crisscrossing chains in the
spinel lattice, as in the checkerboard lattice.

The implication of the absence of the Peierls instability for the case
of $\frac{1}{2}$ an electron per atom in the checkerboard lattice
within one-electron theory is less clear. One possible implication is
that our results for the 2D checkerboard lattice are irrelevant for
the three-dimensional (3D) spinel lattice because of the fundamental
difference between them that has already been pointed out in the
above. In CuIr$_2$S$_4$, the only active orbital following the OO is
the $d_{xy}$ orbital \cite{Khomskii05a}, which has been excluded
within our model. It is thus conceivable that the 1D character of the
active orbitals in CuIr$_2$S$_4$ following OO is much stronger than in
the checkerboard lattice, and this is what drives the metal-insulator
transition in the real system. It is, however, equally likely that the
3D interactions between the $d_{xy}$-based chains are as strong as the
2D interactions in the checkerboard lattice (recall, for example, that
commensurate periodicity for independent 1D chains requires complete
OO, see above). In this case our null result for the uncorrelated
checkerboard lattice would imply non-negligible contribution of e-e
interaction to the metal-insulator transitions in CuIr$_2$S$_4$ and
LiRh$_2$O$_4$. Further theoretical work based on the 3D pyrochlore
lattice as well as experimental work that determines the extent of OO
in the real systems will both be necessary to completely clarify this
issue.

Finally, assuming that e-e interactions play a role, which is subject
to further investigations as discussed above, this raises an
interesting question, viz., what ultimately is the driving force
behind the metal-insulator transitions in CuIr$_2$S$_4$ and
LiRh$_2$O$_4$?  Three of us have argued elsewhere that for carrier
concentration precisely 0.5, there is a strong tendency to form a
paired-electron crystal (PEC), in which there occur pairs of
spin-singlet bonded sites separated by pairs of vacancies
\cite{Li09a}.  This tendency to spin pairing is driven by nearest
neighbor antiferromagnetic (AFM) correlations (as would exist in a
large Hubbard-U system) and is enhanced in the presence of lattice
frustration. Although the original calculations are for the
anisotropic triangular lattice with a single orbital per site, the
same tendency to spin-singlet formation can persist also in the spinel
lattice. If the insulating state in the spinels CuIr$_2$S$_4$ and
LiRh$_2$O$_4$ can be understood as a PEC with spin-singlet pairing
driven by AFM correlations, it may further indicate that e-e
interactions play an important role in superconductivity found in
several structurally-related spinels.  Whether or not e-e interactions
play a role in the observed superconductivity in the spinels
LiTi$_2$O$_4$, CuRh$_2$S$_4$, CuRh$_2$Se$_4$ has remained a lingering
question \cite{Anderson87b,Pickett89a}.  If the insulating state in
this class of materials is indeed a PEC, the superconducting spinels
should perhaps be included among the systems in which
superconductivity is driven not entirely by BCS electron-phonon
coupling.

\section{Acknowledgments}

 Work at Mississippi State University and the University of Arizona
 was supported by the US Department of Energy grant DE-FG02-06ER46315.
 Collaboration between the U.S. teams and the S. N. Bose National
 Centre was supported by a travel grant from the Indo-US Science and
 Technology Forum grant JC/54/2007/``Correlated electrons in
 materials''. The authors gratefully acknowledge discussion with
 D. D. Sarma. SS thanks CSIR for financial support.


\begin{thebibliography}{33}
\bibitem[{foo()}]{footnote}
\bibinfo{note}{We have confirmed this by performing the same calculations with
  degenerate \protect{$p_x$} and \protect{$p_y$} orbitals in which case the
  bonding along the chain directions is \protect{$\sigma$}.}
\end{thebibliography}
\end{document}